\documentstyle[prl,aps,multicol,epsf]{revtex}
\begin{document}
\draft
\title{Field-dependence of
the Magnetic Relaxation in Mn$_{12}$-Acetate:\\
A New Form of Spectroscopy}
\author{Yicheng Zhong and M. P. Sarachik}
\address{Physics Department, City College of the City University of New York,
New York, NY 10031}
\author{Jae Yoo and D. N. Hendrickson}
\address{Department of Chemistry and
Biochemistry, University of California at San Diego, La Jolla, CA 92093}
\date{\today}
\maketitle \begin{abstract}
We report point by point measurements below the blocking temperature of 
the magnetic relaxation of Mn$_{12}$-acetate as a function of
magnetic field applied along the easy axis of magnetization.  Unexpectedly 
complex structure is observed which we attribute to the effect of 
higher-order terms of the spin Hamiltonianon on the 
tunneling process.

\end{abstract}
\vspace {6mm}
\begin{multicols}{2}

The high-spin molecular magnets Mn$_{12}$-acetate (Mn$_{12}$) and Fe$_{8}$ have
received considerable attention as model systems for the study of Quantum
Tunneling of Magnetization (QTM). With a total spin $S=10$ and a large
uniaxial
magnetic anisotropy\cite{sessoli}, Mn$_{12}$ molecules crystallize into a
body-centered tetragonal lattice where the c-axis is along the molecules' easy
axis.  Strong anisotropy breaks each molecule's 21-fold zero-field
degeneracy and results in a double-well potential with energy levels that
correspond to different projections of the spin along the easy axis.  
Strong evidence for QTM in Mn$_{12}$ was obtained in
the form of a series of steps at regular intervals of magnetic field in the
hysteresis loops of Mn$_{12}$ below the blocking temperature of
$\approx 3$ K\cite{friedman1,thomas,hernandez}.  Based on the observation that
the relaxation follows an Arrh\'{e}nius law above 2
K\cite{sessoli,luis},
this process has been attributed to thermally-assisted
tunneling\cite{friedman1,novak_sossli,garanin1}: at these
intermediate
temperatures the
magnetization is thermally activated to some level (or group of adjacent
levels) near the top of the anistropy barrier where the tunneling from
metastable to stable potential wells proceeds. Similar steps in the hysteresis
loops have been observed in Fe$_8$\cite{sangregorio}, where
the quantum mechanical nature of the relaxation process has been established
unambiguously through experiments demonstrating phase interference and parity
effects in response to externally applied transverse fields
\cite{wernsdorfersessoli}.

Measurements of steps in the hysteresis loops do not provide the resolution or
control necessary for a detailed study.  In measurements of the
hysteresis, the external magnetic field is swept continuously at some
predetermined rate so that the magnetic response is generally probed only on
short time scales.  For reasons that are only partially understood, the
initial
response of the magnetization in Mn$_{12}$ is not
characteristic
of the long-time behavior of the relaxation.  We note also that there is a
time-varying internal field associated with the time-varying sample
magnetization which must be added to the externally applied field to obtain
the total field ${\bf H}_{total} = {\bf H} + \alpha (4\pi {\bf M)}$, where $\alpha$ is a 
constant between $0$ and $1$, depending on sample shape and size.  The transient
conditions of
a hysteresis measurement make it particularly difficult to take this effect
into
account.

In this paper we report point-by-point measurements of the magnetic relaxation
of Mn$_{12}$ in fixed external magnetic fields applied along the uniaxial
anisotropy direction which reveal
that the resonances corresponding to steps in the hysteresis
loop consist of rich and quite complex substructures of resolved peaks.  We 
attribute the observed splittings to the higher order longitudinal anisotropy term 
$AS_z^4$ in the
spin Hamiltonian, a small anharmonic contribution that causes
different pairs of energy levels on opposite side of the anisotropy barrier
to be in resonance at slightly different values of magnetic field.  Odd-even 
asymmetry between steps indicates 
that higher order transverse anisotropy also plays a role in Mn$_{12}$.

Millimeter size single crystals of Mn$_{12}$ were prepared according to
the published procedure\cite{lis}.  Parallel and perpendicular components
of the DC magnetization were
obtained with a Quantum Design MPMS-$5$ magnetometer.
A sample rotator was used to align the crystal c-axis parallel to the
magnetic field in increments of
$0.1^{\circ}$.  The
ratio between perpendicular and parallel SQUID readings was typically
$10^{-3}$,
corresponding to $20$ Oe at 2 Tesla.
Cooling the sample in zero
magnetic field before each sweep, full curves of the magnetic relaxation
were recorded
as a function of time in a series of fixed externally applied longitudinal
magnetic fields in the temperature range between $1.8$ K and $2.6$ K.
Following our previous method
of data analysis\cite{friedmanthesis}, we obtained the characteristic rate by
fitting the long
time relaxation to a single exponential.
The overall change in magnetization is small after about $2000$ s,
corresponding to a negligible change in the average internal field $\alpha (4\pi M)$
compared to the sizable (fixed) external field $H$\cite{zhongthesis}.
\vbox{
\vspace{0.2in}
\hbox{
\hspace{-0.2in} 
\epsfxsize 3.3in \epsfbox{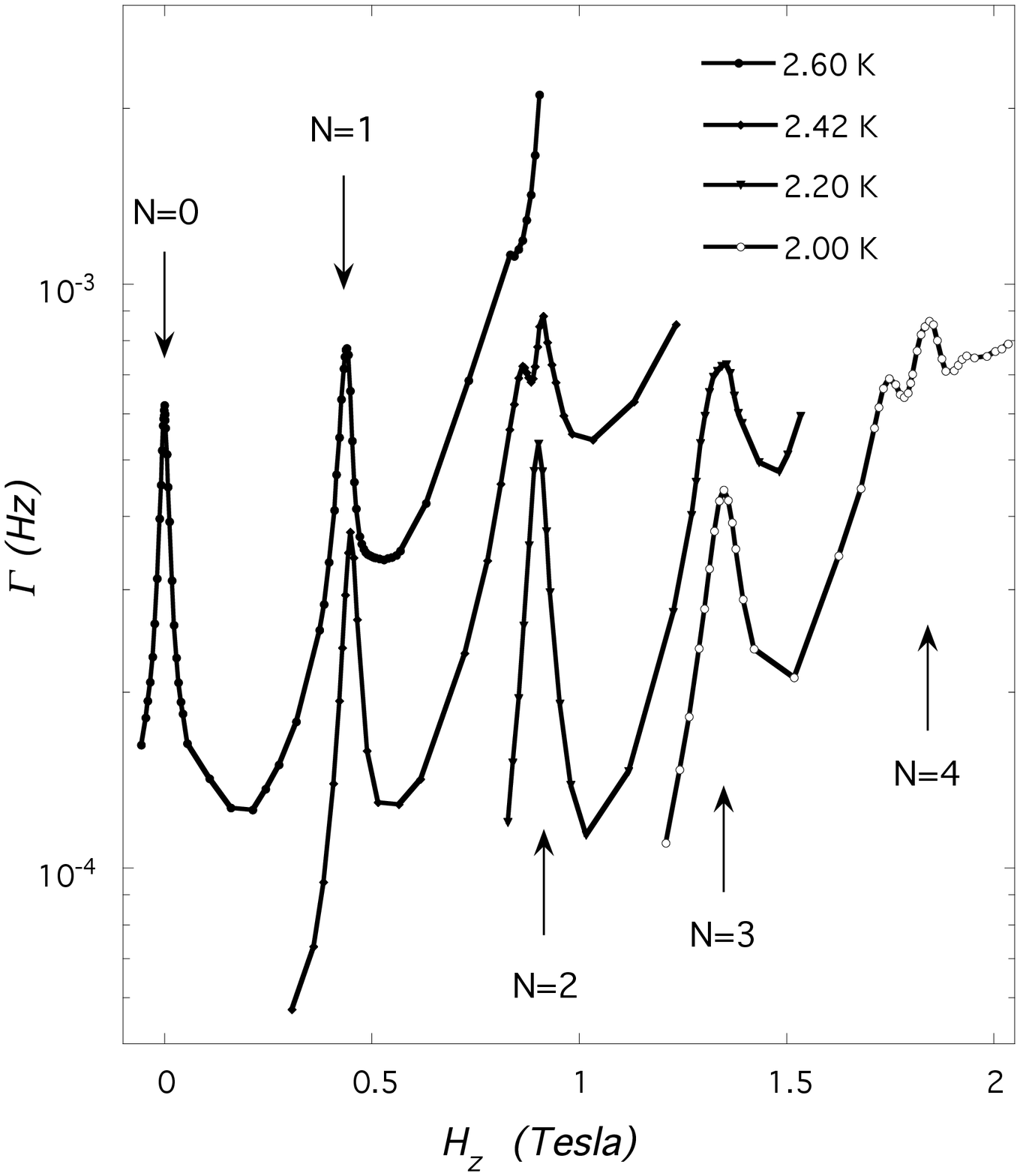} 
}
}
\refstepcounter{figure}
\parbox[b]{3.3in}{\baselineskip=12pt \egtrm FIG.~\thefigure.
Relaxation rate as a function of magnetic field $H_{total}$ applied along the easy 
axis of a single crystal of Mn$_{12}$.  Here $H_{total} = H + \alpha (4\pi M)$ with 
$\alpha = 0.57$.
\vspace{0.10in}
}
\label{1}

The relaxation rates at several temperatures between $1.8$ K and $2.6$ K are
plotted in Figs. 1 and 2 as a function
of the total
longitudinal magnetic field $H_{total} = H + \alpha (4\pi M)$; here the first term is the
externally applied field $H$ and the second term is the contribution
due to the sample's magnetization.  We used $\alpha=0.57$ 
obtained from earlier measurements by Friedman\cite {friedmanthesis} 
on similar samples.

Several observations can be
readily
made from the data.
(1) The main features of the curves confirm the results
obtained from hysteresis measurements: there is faster relaxation at
approximately regularly spaced values of the magnetic field.  Within the 
temperature range of our experiments 
five peaks or resonances are observed, labeled by ``step
numbers''
$N= 0, 1, 2, 3, 4$.  (2) Some of the resonance peaks corresponding to the
``steps'' in the hysteresis loops exhibit additional internal structure.  Most
noticeably, the $N=2$ and $N=4$ peaks split into two or more small peaks at
the higher temperatures.  The $N=1$ and $N=3$ resonances have much
simpler structure.  However, a careful examination of the $N=3$ resonance
at $T=2.20$ K reveals that it also consists of two closely spaced
peaks.  The $N=0$ peak exhibits no apparent structure
at $2.60$ K, or at other
temperatures measured earlier\cite{friedman2}.  (3) Unlike most spectra, which
generally become sharper and more detailed as the temperature is reduced, it is 
interesting
to note that the structure is more complex at higher temperatures.  This is
particularly clear for the $N=4$ resonance where three maxima are clearly seen
at $2.00$ K, while
only one major peak (probably with a small right shoulder) is present at
$1.80$ K. (4) The separation of approximately $0.10$ T between two neighboring
peaks of the $N=4$ resonance is twice as large as
in the case of the $N=2$ resonance, where it is about $0.05$ T (see the
$T=2.00$ K and $2.47$ K curves, respectively).

\vbox{
\vspace{0.2in}
\hbox{
\hspace{-0.2in} 
\epsfxsize 3.3in \epsfbox{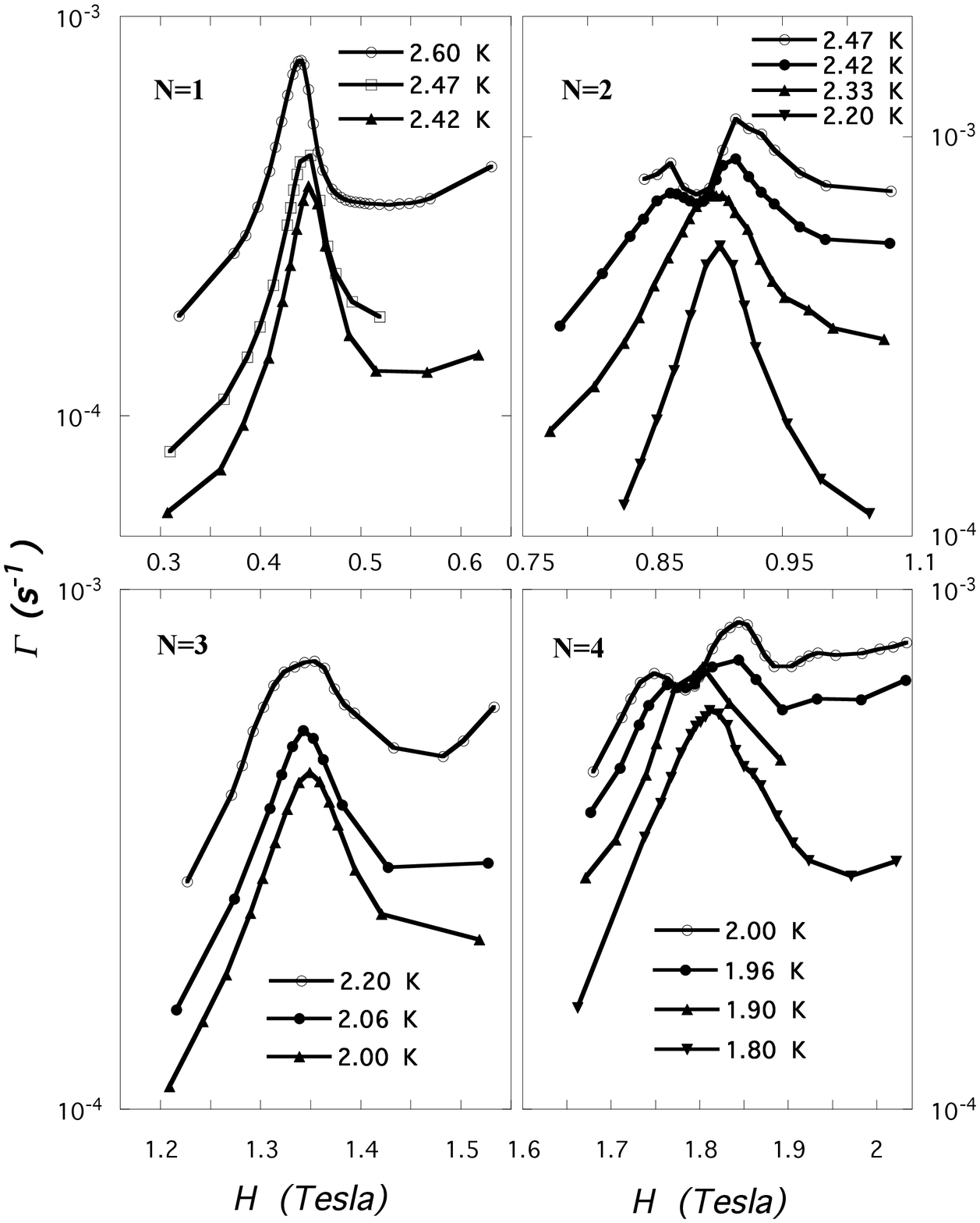} 
}
}
\refstepcounter{figure}
\parbox[b]{3.3in}{\baselineskip=12pt \egtrm FIG.~\thefigure.
Relaxation rate at several temperatures as a function of longitudinal
magnetic field, displaying detailed structure within the $N=1$, $N=2$, $N=3$,
and $N=4$ resonances.
\vspace{0.10in}
}
\label{1}

Hysteresis and earlier relaxation studies of Mn$_{12}$ indicated relatively
simple structure in the form of a series of clean steps, each due to energy
coincidences of pairs of levels on opposite sides of the anisotropy barrier 
corresponding to different spin projections.  In contrast, our detailed studies
reveal that each resonance corresponding to a step displays rather
complex structure.  We suggest that these spectroscopic features are due to the 
non-simultaneous crossing of different pairs of magnetic sublevels, 
as explained below.

Up to fourth
order terms, the spin Hamiltonian that describes Mn$_{12}$ can be written as:
\begin{equation}
{\cal H}= -DS_z^2-\mu_B{\bf H}\cdot {\bf g}\cdot {\bf S}-AS_z^4+C(S_+^4+S_-^4)
\end{equation}
where $D=0.548(3)$ K is the anisotropy constant, the second term represents
the
Zeeman energy, and the remaining are higher-order terms in the crystalline
anisotropy allowed by tetragonal symmetry. A level in the metastable potential
well with magnetic quantum
number $m$ comes into resonance with level
$m'$ in the
stable well at a longitudinal magnetic field $H_z$ given by:
\begin{equation} H_z=N\frac{D}{g_z  \mu_B}\left[1 + \frac{A}{D}\left(m^2 +
m'^2\right)\right]. \label{equation}
\end{equation}
Electron paramagnetic resonance (EPR)\cite{barra} and inelastic neutron
scattering experiments in Mn$_{12}$\cite{zhong,mirebeau} have shown
there is a sizable fourth order term, $AS_z^4$, with $A=1.173(4) \times 10^{-3}$
K.  As a consequence,
different pairs of levels
$m, m'$ on opposite sides of the anisotropy barrier come into resonance at
slightly different magnetic fields for the same step number
$N = \mid m+m' \mid$.  Note that higher fields are necessary to bring
lower-energy pairs
(bigger $|m|$) into resonance, an effect that is more pronounced for peaks
with higher numbered steps (due to the multiplicative factor $N$ in
Eq. (2))\cite{kent}.  Using the parameters $D$, $A$ and $g_z$ obtained by
inelastic neutron scattering\cite{mirebeau} and EPR measurements\cite{barra},
the magnetic fields at which pairs of levels $m, m'$ come into resonance
calculated from Eq. (2) are listed in Table I.

Several features of the data are consistent with this model.  In 
agreement with Eq. (2), the separation between neighboring peaks is generally 
larger within the higher number steps: the splitting 
for $N=4$ is twice that for $N=2$, there is no substantial splitting (or shift) 
for the $N=1$ resonance, and none at all for $N=0$.  Also, since the tunneling 
is thermally assisted in this temperature regime, one might expect that spin 
reversal through resonant levels near the top of the potential well will become 
more active, introducing additional channels for tunneling as the temperature 
is raised.  This could account for the fact that more complex structure is observed 
at higher temperatures.  Quantitative comparison of the measured peak positions
and the calculated resonant fields listed in Table I requires a reliable 
determination  
of the value of $\alpha$ used to calculate the total field 
$H_{total} = H + \alpha (4\pi M)$.  Using $\alpha= 0.57$ found in earlier 
experiments on similar material\cite{friedmanthesis}, we suggest that 
the $N=1$ peak centered at about $0.437$ T at $2.60$ K is 
associated with tunneling from $m=-3$ and/or $-4$ in the metastable well, while 
at the lower temperatures of $2.47$ K and $2.42$ K, the $N=1$ maximum observed at 
$0.446$ T is very close to the field $0.443$ T listed in Table I for tunneling from 
$m=-4$ to $m'=3$.

Other features of the data are more difficult to understand and warrant 
further discussion.  One is that
there appears to be considerably more structure in even ($N=2,4$) than in odd
($N=1,3$) resonances within the range of field and temperature of our
measurements.  It is puzzling, for example, that the $N=3$ peak shows much
less structure than the $N=2$ resonance (note the multiplicative factor
$N$ of Eq. (1)).  A second interesting feature is that the splitting within the 
$N=2$ and $N=4$ steps are too large to
correspond to the difference in magnetic fields for two immediately
neighboring level crossings ({\it i. e.} tunneling from $m$ and
($m-1$)).  Instead, they correspond more closely to the field difference for every
{\it other} level crossing ({\it i.e.} tunneling from $m$ and ($m-2$)).
In constrast,
the $N=3$ resonance has negligible
splitting and its two closely-merged peaks are probably due to immediately
neighboring level crossings.  These effects suggest that 
higher-order transverse anisotropy is important in the tunneling process.

Two symmetry-breaking terms that could be responsible for 
tunneling have been considered: (1) a transverse magnetic 
field (of hyperfine/dipolar origin, or externally applied); and (2) an 
anisotropy term of the form $C(S_+^4+S_-^4)$, the lowest order allowed by the 
tetragonal symmetry of Mn$_{12}$.  Our data presents a clear enigma.  On the one 
hand, the odd-even asymmetry between steps and 
level-skipping within even-numbered steps both suggest that transverse anisotropy 
plays an important role in the tunneling.  On the other hand, 
the odd and even-numbered resonances have comparable amplitudes, 
albeit at different temperatures.  This implies that the tunneling is due to 
transverse magnetic fields.  A simple 
model\cite{calculation} calculation for the tunneling rates that includes both 
symmetry-breaking 
terms does not fit the experimental results for any values of the parameters: 
the odd-even asymmetry requires that transverse anisotropy dominate, 
while the comparable amplitudes require that a transverse field be the dominant 
symmetry-breaking term.  Which mechanism is 
responsible for tunneling in this material is an interesting question that has yet 
to be resolved.

Leuenberger and Loss\cite{LL} and Pohjola and Schoeller\cite{pohjola} recently
presented a comprehensive theory of the magnetization relaxation in
Mn$_{12}$ in the thermally assisted tunneling regime. The relaxation rate as a 
function of longitudinal magnetic field was calculated using a Hamiltonian that
includes spin-phonon interactions, quartic magnetic anisotropy and a weak 
transverse field. They obtained satellite peaks qualitatively similar to those 
shown in Figs. 2 and 3 whose amplitude and width they investigated\cite{LL} for 
various angles between the sample easy axis and the field direction. 
Quantitative agreement between their theory and our experimental observations 
requires a substantial transverse magnetic field, due perhaps to mosaic spread or 
unintentional misalignment of the sample with respect to the magnetic 
field direction.  However, the mosaic spread of approximately $0.4^{\circ}$ 
\cite{robinson} found for Mn-12 crystals is too small to account for our 
observations, and the degree of 
misalignment required to give a sufficiently large transverse component is well 
outside our experimental margin of uncertainty and would yield different results 
each time a sample is mounted and aligned.

In closing, we note that the point-by-point relaxation measurements
presented in this paper represent a novel form of spectroscopy.
``Spectra'' obtained from such measurements allow detailed studies of
the tunneling process, and provide important information regarding the
dominant
tunneling paths at different temperatures and magnetic fields.

We are grateful to M. Leuenberger, D. Loss and T. Pohjola for numerous discussions 
and detailed comparison between theory and experiment.  We thank Jonathan 
Friedman and Eugene Chudnovsky for their continued interest 
and for valuable comments on our manuscript, and Jonathan Friedman for 
suggesting this experiment.  This work was supported by NSF grant DMR-9704309.

\begin{table}
TABLE I: From Eq. (2), calculated values of magnetic fields at which a
level $m$ in the
metastable well is in resonance
with level $m'$ in the stable well.
Values for parameters $D=0.548(3)$ K, $A=1.173(4) \times 10^{-3}$ K,
and $g_z=1.94(1)$ are taken from recent neutron scattering and
EPR experiments.
\label{table1}
\end{table}

\begin{tabular}{|c|c|c|c|c|c|c|c|c|}	\hline
 	 & \multicolumn{2}{c|}{$N=1$} & \multicolumn{2}{c|}{$N=2$} &
\multicolumn{2}{c|}{$N=3$}
 	 & \multicolumn{2}{c|}{$N=4$} \\ \hline\hline
 $m$  & $m'$ & H (T) & $m'$ & H (T) & $m'$ & H (T) & $m'$ & H (T) \\ \hline
 $-1$  & $0$ & $0.422$ & --- & --- & --- & --- & --- & --- \\ \hline
 $-2$  & $1$ & $0.425$ & $0$ & $0.848$ & $-1$ & $1.275$ & --- & --- \\ \hline
 $-3$  & $2$ & $0.432$ & $1$ & $0.859$ & $0$ & $1.286$ & $-1$ & $1.719$ \\
\hline
 $-4$  & $3$ & $0.443$ & $2$ & $0.877$ & $1$ & $1.308$ & $0$ & $1.740$ \\
\hline
 $-5$  & $4$ & $0.458$ & $3$ & $0.902$ & $2$ & $1.340$ & $1$ & $1.776$ \\
\hline
 $-6$  & $5$ & $0.476$ & $4$ & $0.935$ & $3$ & $1.383$ & $2$ & $1.827$ \\
\hline
 $-7$  & $6$ & $0.497$ & $5$ & $0.974$ & $4$ & $1.437$ & $3$ & $1.891$ \\
\hline
 $-8$  & $7$ & $0.522$ & $6$ & $1.021$ & $5$ & $1.502$ & $4$ & $1.971$ \\
\hline
 $-9$  & $8$ & $0.551$ & $7$ & $1.075$ & $6$ & $1.578$ & $5$ & $2.064$ \\
\hline
 $-10$  & $9$ & $0.584$ & $8$ & $1.137$ & $7$ & $1.664$ & $6$ & $2.172$ \\
\hline
\end{tabular}

\end{multicols}

\begin{references}
\bibitem{calculation} The tunneling probabilities were calculated from 
the thermal population factors and the tunnel splittings obtained by 
diagonalizing the $21 \times 21$ matrix for the spin Hamiltonian of Eq. (1) for 
Mn$_{12}$-acetate.  The observed tunneling rates could not be reproduced for 
any choice for the size of the transverse terms 
$g\mu_B H_x$ and $C(S_+^4+S_-^4)$.
\end{references}
\end{document}